\documentclass{webofc}
\usepackage[varg]{txfonts}

\begin{document}

\title{Enhancements of high order cumulants across the 1st order phase
transition boundary}
\author{\firstname{Lijia} \lastname{Jiang}\inst{1,2}\fnsep\thanks{%
jiang@fias.uni-frankfurt.de} \and \firstname{Shanjin} \lastname{Wu}\inst{2}%
\fnsep\thanks{%
shanjinwu2014@pku.edu.cn} \and \firstname{Huichao} \lastname{Song}%
\inst{2,3,4}\fnsep\thanks{%
huichaosong@pku.edu.cn} }

\institute{ Frankfurt Institute for Advanced Studies,Ruth Moufang Strasse 1, D-60438,
Frankfurt am Main, Germany
\and
 Department of Physics and State Key Laboratory of Nuclear Physics and
Technology, Peking University, Beijing 100871, China
\and
Collaborative Innovation Center of Quantum Matter, Beijing 100871, China
\and
Center for High Energy Physics, Peking University, Beijing 100871, China          }

\abstract{ In this proceeding, we investigate the dynamical evolution of the $\sigma$ field
with a trajectory across the 1st order phase transition boundary, using the Langevin equation
from the linear sigma model.
We find the high order cumulants of the $\sigma$ field are largely enhanced during the dynamical
evolution, compared with the equilibrium values, due to the
 supercooling effect of the first order phase transition. }
\maketitle
\section{Introduction}

\quad The STAR Beam Energy Scan (BES) program aims at searching and locating
the QCD critical point. It was proposed that the fluctuations of net baryon number
and net protons are largely enhanced  in the
critical region  due to the  increased
correlation length~\cite{Stephanov:2008qz, Stephanov:2011pb}. In experiment, large deviations of high order cumulants
and cumulants ratios of net protons from the Poisson baselines have been observed at the collision energies
below 40 GeV~\cite{Aggarwal:2010wy, Adamczyk:2013dal, Luo:2015ewa}, indicating the potential of discovering the critical point.

Many past research has studied the static critical fluctuations
near the critical point, which could explain the acceptance dependence of the net proton's fluctuations,
but could not qualitatively describe all the cumulants measured in experiment~\cite{Ling:2015yau,Jiang:2015hri,Jiang:2015cnt}.
Recently, the investigations on the dynamical critical fluctuations based on the Fokker-Plank equation and Langevin dynamics
have shown that critical slowing-down effects play important roles in the critical regime, where the signs of high order dynamical cumulants of the $\sigma$ field are even different from the corresponding equilibrium
values~\cite{Mukherjee:2015swa, Jiang:2017mji}.

In this proceeding, we  focus on studying the
dynamical evolution of the $\sigma$'s cumulants with the 1st order phase transition
scenario, using the Langevin dynamics within the framework of the linear sigma model.
We will show that the high order cumulants of the $\sigma$ field are largely enhanced during the dynamical
evolution when compared with the equilibrium values, due to the
supercooling effect across the first order phase transition boundary.

\begin{figure*}[tbp]
\center
\includegraphics[width=5.6 in, height=1.6 in]{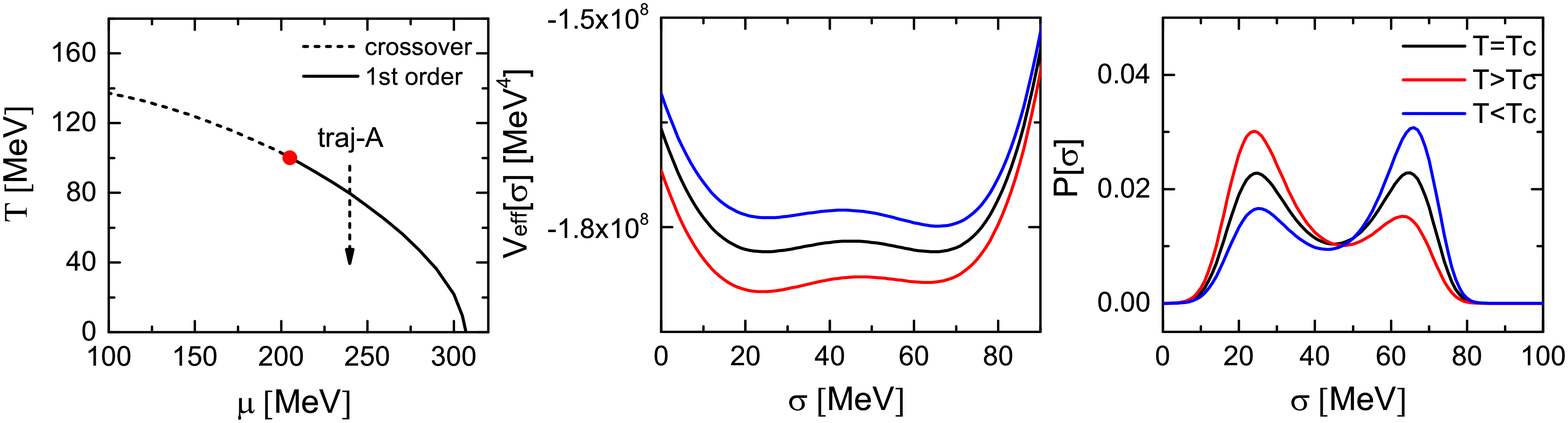}
\caption{ Left panel: the phase diagram of the linear sigma model. The middle and right panel:
the effective potential $V_{eff}(\sigma)$ and probability distribution function $P[\sigma]$ at
different temperature $T>T_c$, $T=T_c$ and $T<T_c$, along a chosen trajectory with $\protect\mu = 240$ MeV.}
\label{fig1}
\end{figure*}

\section{The formulism}

\quad Within the framework of the linear sigma model with constituent quarks, we solve the time evolution of the $\sigma$ field based on the Langevin equation. The effective potential of this model gives a
 phase diagram  with cross-over, the first order phase transition as well as a critical
point~\cite{Jungnickel:1995fp,Skokov:2010sf,Nahrgang:2011mg}.
The long wavelength mode of the $\sigma $ field is the
order parameter of the chiral phase transition, and its mass vanishes at the
critical point. According to the classification of \cite{Hohenberg:2017}, the dynamics of the $\sigma$ field
in the critical regime belongs to model A, which only evolves the non-conserved order parameter field.
The Langevin equation of the $\sigma $ field is written as:
\begin{equation}
\partial ^{\mu }\partial _{\mu }\sigma \left( t,x\right) +\eta \partial
_{t}\sigma \left( t,x\right) +\frac{\delta V_{eff}\left( \sigma \right) }{%
\delta \sigma }=\xi \left( t,x\right) ,
\label{langevin}
\end{equation}%
which is a semi-classical equation. The damping $\eta $ and
noise $\xi \left( t,x\right)$ term in the above equation satisfy the fluctuation-dissipation
theorem: $
\left\langle \xi \left( t\right) \xi \left( t^{\prime }\right) \right\rangle
\sim \eta \delta \left( t-t^{\prime }\right) $.
The equilibrium thermodynamical potential is obtained by integrating out the
thermal quarks:
\begin{equation}
V_{eff}\left( \sigma \right) =U\left( \sigma \right) +\Omega _{\bar{q}%
q}\left( \sigma \right) =\frac{\lambda ^{2}}{4}\left( \sigma
^{2}-v^{2}\right) ^{2}-h_{q}\sigma -U_{0}+\Omega _{\bar{q}q}\left( \sigma
\right) .\qquad \quad
\end{equation}
Here $U\left( \sigma \right) $ represents the vacuum potential of the $\sigma $
field, $\Omega _{\bar{q}q}\left( \sigma \right) =-d_{q}\int \frac{d^{3}p}{%
\left( 2\pi \right) ^{3}}\{\omega+T\ln [1+e^{-\left( \omega-\mu \right) /T}]+T\ln
[1+e^{-\left( \omega+\mu \right) /T}]\}$ comes from the contribution of thermal quarks,
the particle energy $\omega=\sqrt{p^2+M^2}$, and the effective mass of quarks $M(\sigma)=m_0+g\sigma$,
with the current quark mass $m_0=5.5$ MeV, and coupling $g=3.3$.
The parameters $h_{q}, \nu, \lambda, U_{0}$ in the effective potential are
determined by the properties of vacuum hadrons. Due to the spontaneously broken of chiral symmetry in vacuum, the vacuum expectation of $\sigma$ field is $\left\langle \sigma \right\rangle =f_{\pi }=93$ MeV. $h_{q}$ is the symmetry explicitly broken term, $h_{q}=f_{\pi }m_{\pi }^{2}$ with $m_{\pi }=138$ MeV. The parameter $\nu$ is determined by $\nu ^{2}=f_{\pi
}^{2}-m_{\pi }^{2}/\lambda ^{2},$ the in vacuum mass of $\sigma$ is $m_{\sigma
}\sim 600$ MeV by setting $\lambda ^{2}=20.$ The zero-point energy $%
U_{0}=m_{\pi }^{4}/\left( 4\lambda ^{2}\right) -f_{\pi }^{2}m_{\pi }^{2}.$
Note that we have neglected the fluctuations of $\vec{\pi}$,
since its mass is finite in the critical regime.

From the thermodynamic potential in eq.\,(2), we depict the phase diagram on the $\mu$-$T$ plane
with a critical point located at $%
\left(\mu _{c}, T_{c}\right) =\left( 205,100.2\right) $ MeV, as shown in Fig.~\ref{fig1} (left).
The middle and right panel of  Fig.~\ref{fig1} present
the effective potential and probability distribution function of the $\sigma$ field $\left( P\left[ \sigma \left( \mathbf{x}\right) \right]
\sim \exp \left( -E\left( \sigma \right) /T\right)
~\text{with}~ E\left( \sigma
\right) =\int d^{3}x\left[ \frac{1}{2}\left( \nabla \sigma \left( x\right)
\right) ^{2}+V_{eff}\left( \sigma \left( x\right) \right) \right] \right)$ at
different temperature $T>T_c$, $T=T_c$ and $T<T_c$, along a chosen trajectory with $\protect\mu = 240$ MeV.

For the numerical implementation of the Langevin dynamics, we need to input  the initial
configurations of the $\sigma $ field to start the simulations.  Here,  we construct the event-by-event initial
conditions according to the probability function $P\left[ \sigma \left( \mathbf{x}\right) \right]$.
In addition, the simulations of
the Langevin equation requires to input the changing temperature and chemical
potential profiles $T\left( t,x,y,z\right) $ and $\mu \left( t,x,y,z\right)$ from the heat bath.  For simplicity, we assume the system is isothermal and evolves along
a trajectory across the first order phase transition boundary
with a fixed chemical potential, and the temperature decreases
in a Hubble like way: $\frac{T\left( t\right) }{T_{0}}=\left( \frac{t}{t_{0}}%
\right) ^{-0.45}.$

From the dynamical evolution of the $\sigma $ field, one can obtain the spatial
information of $\sigma (x)$ at each time step for each event, and then sum
all the event configurations to obtain the associated moments:
$\mu _{n}^{\prime }=\langle \sigma ^{n}\rangle
=\frac{\int d\sigma \sigma ^{n}P\left[ \sigma \right] }{\int d\sigma P\left[
\sigma \right] }$, where $\sigma =\int d^{3}x\sigma \left( \mathbf{x}\right)
$.  The cumulants of the
$\sigma$ field can be calculated from these various moments with:%
\begin{eqnarray}
C_{1} =\mu _{1}^{\prime },\;C_{2}=\mu _{2}^{\prime }-\mu _{1}^{\prime
2},\;
C_{3}=\mu _{3}^{\prime }-3\mu _{2}^{\prime }\mu _{1}^{\prime }+2\mu
_{1}^{\prime 3},\;
C_{4} =\mu _{4}^{\prime }-4\mu _{3}^{\prime }\mu _{1}^{\prime }-3\mu
_{2}^{\prime 2}+12\mu _{2}^{\prime }\mu _{1}^{\prime 2}-6\mu _{1}^{\prime 4}.
\end{eqnarray}

\section{Numerical results}
\begin{figure*}[tbp]
\center
\includegraphics[width=3.7 in]{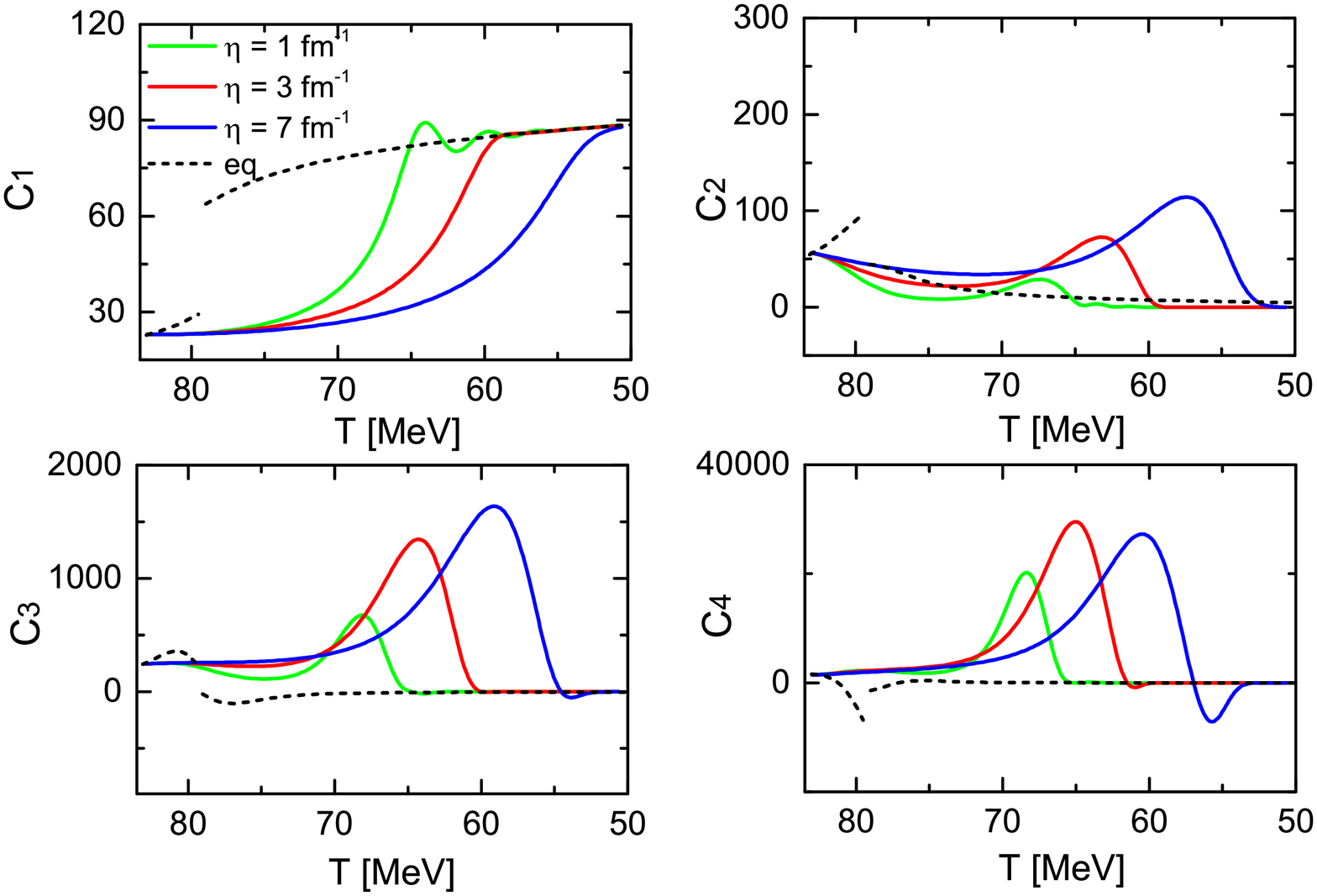}
\caption{Dynamical evolution of the $\sigma$'s cumulants along trajectory-A across the first order
phase transition with $\mu = 240 $ MeV. The colored lines
represent the dynamical cumulants with different damping coefficients.
The black dotted lines are the corresponding equilibrium cumulants.}
\label{fig3}
\end{figure*}

\quad In this proceeding, we focus on investigating the evolution of the $\sigma$ field along a trajectory
across the first-order phase transition boundary (as shown in Fig.\,1
traj-A). Fig.\,2 shows the time evolution of dynamical cumulants of the $\sigma $ field
along traj-A, with different damping coefficients (solid lines with different colors). The black dotted lines represent the
equilibrium cumulants of the $\sigma$ field at fixed temperature and chemical potential
along the trajectory of the evolving heat bath, obtained from the perturbative calculations around the global minimum~\cite{Stephanov:2008qz}.

As shown in Fig.\,2, the magnitude of damping coefficients directly influence the dynamical
evolution of $\sigma$ field. A larger damping coefficient  $\eta$ leads to a slower
evolution of the $\sigma $ field. Besides, the  evolution with a larger damping
coefficient tends to develop larger fluctuations at late evolution time.
For all three chosen damping coefficients, we find clear memory effects. When the
system goes below $T_c$, the high order cumulants, especially $C_3$ and $%
C_4$ can memorize the sign of the earlier stage during the evolution.

On the other hand, the dynamical cumulants of the $\sigma $ field are largely enhanced compared
with the equilibrium values, after the system evolves across the first order phase
transition boundary. As shown in Fig.\,\ref{fig1} (middle and right),  the effective potential and corresponding probability distribution function shows double well (peak) structures near the first order phase transition above and below $T_c$.
Due to the supercooling effect,
the $\sigma$ field tends to distribute in both minima of the potential, which largely increases the magnitude of
high order cumulants.

\section{Summary}

\quad In this proceeding,  we numerically simulate the dynamical evolution of the $\sigma$ field
with a trajectory across the 1st order phase transition boundary, using the event-by-event
simulations of the Langevin dynamics.  The dynamical cumulants
present clear memory effects compared with the equilibrium ones, where $C_3$ and $C_4$ can memorize
the signs of the early stage during the dynamical evolution. Besides, the dynamical critical
fluctuations are largely enhanced after the system across
the 1st order phase transition boundary, due to the associated suppercooling effect.

\end{document}